# Excitation efficiency and limitations of the luminescence of $Eu^{3+}$ ions in GaN

D. Timmerman[1], B. Mitchell[2], S. Ichikawa[1], J. Tatebayashi[1], M. Ashida[3] and Y. Fujiwara[1]

[1] Graduate School of Engineering, Osaka University, 2-1 Yamadaoka, Suita, Osaka 565-0871, Japan.

[2] Department of Physics, West Chester University, West Chester, PA, 19383, USA

[3] Graduate School of Engineering Science, Osaka University, 1-3 Machikaneyama, Toyonaka, Osaka 560-8531, Japan

The excitation efficiency and external luminescence quantum efficiency of trivalent $Eu^{3+}$ ions doped into gallium nitride (GaN) was studied under optical and electrical excitation. For small pump fluences it was found that the excitation of $Eu^{3+}$ ions is limited by an efficient carrier trap that competes in the energy transfer from the host material. For large pump fluences the limited number of high-efficiency $Eu^{3+}$ sites, and the small excitation cross-section of the majority $Eu^{3+}$ site, limit the quantum efficiency. At low temperatures under optimal excitation conditions, the external luminescence quantum efficiency reached a value of 46%. These results show the high potential for this material as an efficient light emitter, and demonstrates the importance of the excitation conditions on the light output efficiency.

**Introduction**

Eu-doped GaN shows great promise for its application as a red-light emitter with temperature insensitive, sharp and stable emission based on the well-developed GaN platform. Current injection is relatively easy in these materials, and high optical output powers of device structures have been shown[1,2]. However, past research has shown that the excitation mechanism of the Eu-related emitting centers is diverse and their optical activity is determined by various factors. The origin of this variation lies in the existence of multiple incorporation sites of the Eu ions in the GaN host. The variation of their local environments gives rise to small variations in the excitation and emission wavelength, which allows the local environments to be distinguished[3–7]. The majority center in GaN:Eu, which is referred to as OMVPE4 (Eu1), shows emission around 622 nm and comprises about 90% of the total $Eu^{3+}$ ions. Two other important centers known as OMVPE7 (Eu2) and OMVPE8 (Eu2*), which have been shown to differ only in charge state for the same defect configuration[8], show the most pronounced emission under current injection. It is noted that, at room temperature, it is not possible to distinguish OMVPE4 and OMVPE7 (~622 nm), while the peak of OMVPE8 (~618 nm) is easy to isolate. These two sites comprise < 10% percent of the $Eu^{3+}$ ions, and will be addressed as the minority sites. In this work, we quantitatively determine the basic excitation properties of the minority and majority sites under optical and electrical excitation. Additionally, the factors limiting the output efficiency are determined. It is shown that while the external quantum efficiency of luminescence can reach values up to about 50%, they are strongly dependent on the excitation conditions. These results encourage the use of these materials in micro-LEDs, where relative low current densities are typically employed.

**Experimental**

For the PL experiments a Pharos (Light conversion, Vilnius, Lithuania) operating at 1 kHz with a ~200 fs pulse width and ~0.2 mJ pulse energy at an output wavelength of 1030 nm, was used to pump an OPA, Orpheus-HP (Light Conversion, Vilnius, Lithuania) to generate excitation pulses with a wavelength of 350 nm. The excitation laser was guided through a circular pinhole in order to obtain a homogenous intensity throughout the excitation spot (top-hat profile) for all optical experiments. Luminescence was spectrally dispersed by a 0.5 spectrometer, SpectraPro HRS-300 (Acton Research Corporation, Acton, USA) and detected with an air-cooled CCD, PIXIS 256 (Princeton Instruments, Trenton, USA). For the determination of absolute PL quantum efficiencies the sample was placed in an integrating sphere (LabSphere). Low temperature measurements were performed with a nitrogen-cooled cryostat.

The samples and devices in this work were grown by OMVPE on (0001) sapphire substrates, and trimethylgallium (TMGa) and ammonia ($NH_3$) were used as the gallium and nitrogen sources, respectively. The reactor pressure was maintained at 100 kPa during growth. For the GaN:Eu layers $EuCp^{pm}_2$ was used as the Eu source. The Eu source and transfer lines were maintained at 125 and

135 °C, respectively. The Eu concentration was determined to be $1 \times 10^{20}$ cm$^{-3}$ by secondary ion mass spectroscopy. The growth temperature of the of the optical active layer was 960 °C, which is the optimal growth temperature for GaN:Eu[9,10]. An LED structure was grown using the same structure as the active layer, which was surrounded by an LED structure similar as in Ref. 11. Ti/Au and Ni/Au contacts were formed by electron beam deposition on the n- and p-type layers, respectively.

**Effective excitation cross-section of photoluminescence**

In order to quantitatively determine the efficiency of excitation of the most important sites for above bandgap pumping, power dependent PL spectra were obtained under pulsed excitation. A laser repetition rate of 1 kHz ensured that all excited Eu$^{3+}$ ions relaxed to the ground state before the next pulse arrived[12]. Figure 1a shows the PL spectra for different pump fluences, for an excitation wavelength of 350 nm, where the spectral designations of the peaks are also indicated. The number of Eu$^{3+}$ ions in the excited state that contribute to PL under laser illumination, $Eu^*$ is governed by simple kinetics:

$$\frac{dEu^*[t]}{dt} = \sigma_{ex}\varphi(Eu_{tot} - Eu^*[t]) - \frac{Eu^*[t]}{\tau} \qquad (1)$$

With $Eu_{tot}$ the total number of excitable Eu$^{3+}$ ions, $\tau$ the relaxation time, $\sigma_{ex}$ the effective excitation cross-section and $\varphi$ the laser photon flux. In the case the laser pulse length is short compared to $\tau$, and the repetition rate is low enough to allow all excited Eu$^{3+}$ ions to relax to the ground state, the second term describing the relaxation can be ignored. By replacing the photon flux with the photon fluence; $f = \varphi \times \Delta t$, with $\Delta t$ the laser pulse width, we can simplify the equation to:

$$\frac{dEu^*[t]}{dt} = \sigma_{ex}f(Eu_{tot} - Eu^*[t]) \qquad (2)$$

The solution of this equation is given by:

$$Eu^* = Eu_{tot}(1 - e^{-\sigma_{ex}f}) \qquad (3)$$

Spectral deconvolution was performed to isolate the peaks originating from the OMVPE8 and OMVPE4/7 sites to gain information about them independently. The pump fluence dependence of OMVPE8 is shown in Fig. 1b, together with a fit of Eq. (3). An effective excitation cross-section of 1.6 x 10$^{-15}$ cm$^2$ is found. As the PL lifetime of all peaks have similar values, it can be assumed that the emission efficiency of the different sites is similar. This allows us to use the saturation value of OMVPE8, together with the relative abundance of the sites, to scale the excited fraction of the OMVPE4/7 peak. In the fluence dependence of OMVPE4/7, two regions can be observed (Fig. 1c). In the low fluence region, there is a contribution of both sites, while OMVPE7 saturates at the same excitation fluence as OMVPE8. For larger fluences only, OMVPE4 will increase its emission intensity. A fit of Eq. (3) on the high fluence part of the peak now gives the effective excitation cross-section of the majority site OMVPE4 with a value of 1.2 x 10$^{-17}$ cm$^2$, slightly more than two orders of magnitude smaller than the high efficiency sites.

**External photoluminescence quantum yield**

More information about the excitation efficiency can be obtained by comparing the excitation fluence and PL intensity under the same excitation conditions as before. By taking the ratio of these two, a relative quantum efficiency of the PL efficiency has been determined. Simultaneously, the absolute quantum efficiency has been determined by an integrating sphere methodology, in order to calibrate the relative values (See supplementary information). The results for a variation of the pump fluence over 6 orders of magnitude is shown in Fig. 2. For small fluences the QE increases as function of the pump fluence, until it reaches a maximum around 0.05 mJ cm$^{-2}$, while for larger pump fluences the QE decreases again. From a linear plot of the low pump fluence regime, it can be observed that the QE quickly rises until about 5 µJ cm$^{-2}$, after which it increases more gradually. This behavior can be explained by an efficient trapping center that competes with the Eu ions, lowering its QE, and reduces its influence when all traps have been filled. The photon fluence for which this center is saturated, is about $10^{13}$ cm$^{-2}$, i.e. the trapping center has an effective cross-section in the order of ~$10^{-13}$ cm$^2$. The most likely candidate for this center is the H1 hole trapping center[13,14]. It is typically found in OMVPE grown GaN and has a large carrier capture cross-section of about $10^{-13}$ cm$^{-2}$, with a typical concentration of a few times $10^{16}$ cm$^{-3}$. The origin of this trap is controversial and has been attributed to gallium vacancies ($V_{Ga}$)[15] or C-related defects ($C_N$)[16]. For completeness it is noted that other native defects are also contributing in this pump fluence region in competition with Eu excitation, but these have typically lower concentrations and/or smaller carrier capture coefficients, and thus a more limited contribution.

In the high pump fluence regime, the PL QE drops as the number of available Eu$^{3+}$ ions for excitation limits the emission after a single pulse. We note that the determination of the PL QE after a short excitation pulse gives a much "cleaner" result when compared to continuous wave (cw) excitation, as the dynamics of the competing carrier traps are less important. If the carrier trap lifetimes are relatively short (compared to the Eu emission) the observed PL QE will generally be lower in cw excitation, as these traps can trap multiple carriers before Eu ions emit and can be re-excited again. This explains why much lower values of the external QE efficiency has been found before under low power cw excitation[17]. Also in that work, under pulsed excitation, a much lower value was found for the PL QE (~0.03) for a photon fluence of 4.6 x $10^{16}$ cm$^{-2}$, which is actually in line with the determined values here.

Under intense cw excitation, which is a common experimental condition that is used for the study of these materials, thermal quenching results in a PL intensity at room temperature that is ~15% the value observed at low temperature[7]. However, this is different under the pulsed excitation conditions used in this study. Figure 3 shows the PL spectrum and PL QE as function of temperature in the range 77-300 K under the excitation conditions where the QE has its maximum. It can be seen that the QE increases relatively continuously up to a value of 0.46. The minority site OMVPE8 is hardly quenched and shows only a small increase in intensity when going to lower temperature, while the intensity from the majority site nearly doubles for low temperature.

**Excitation cross-section of electroluminescence**

Under current injection of GaN:Eu LEDs, the spectrum changes slightly (Fig. 2a), and the ratio of the peaks for high current doesn't change as much as under optical excitation. Since this concerns a cw experiment, the rate equations describing the excitation, and solution of it, are different. The number of $Eu^{3+}$ ions in the excited state that contribute to EL under current injection, $Eu^*$ is governed by the following kinetics:

$$\frac{dEu^*[t]}{dt} = \sigma_{ex}\frac{j}{q}(Eu_{tot} - Eu^*[t]) - \frac{Eu^*[t]}{\tau} \qquad (4)$$

with $Eu_{tot}$ the total number of excitable Eu ions, $\tau$ the relaxation time, $\sigma_{ex}$ the excitation cross-section, $j$ the current density and $q$ the elementary charge. The solution to this equation by:

$$Eu^*[t] = Eu_{tot}\frac{(1-e^{-t(\frac{1}{\tau}+\sigma_{ex}\frac{j}{q})})\sigma_{ex}\tau\frac{j}{q}}{1+\sigma_{ex}\tau\frac{j}{q}}, \qquad (5)$$

which reaches an equilibrium value for long times given by:

$$Eu^* = Eu_{tot}\frac{\sigma_{ex}\tau\frac{j}{q}}{1+\sigma_{ex}\tau\frac{j}{q}} \qquad (6)$$

As with the optical excitation, the peaks have been spectrally deconvoluted to determine their relative intensities. The current density dependence of OMVPE8 is depicted in Fig. 4b, together with a fit of Eq. (6). A value for the excitation cross-section of 3.0 x $10^{-15}$ cm$^2$ was found, which is about a factor of two larger than the effective excitation cross-section under optical excitation. We exploit the relationship between OMVPE7 and OMVPE8 to subtract a rescaled intensity dependence from the OMVPE4/7 peak in order to isolate the OMVPE4 contribution. It can be seen that for a current density over 1 A/cm$^2$ this gives a linear dependence. Following the same rescaling procedure as under optical excitation, the relative fraction of excited majority centers has been determined. Subsequently, the high current density dependence has been fitted with Eq. (6), and resulted in an excitation cross-section of 6 x $10^{-18}$ cm$^2$, which is about a factor of two smaller than under optical excitation.

**Discussion**

Multiple processes compete with radiative deexcitation of the $Eu^{3+}$ ions and decrease the efficiency of *4f-4f* luminescence, most prominently at elevated temperatures[18]. The two most important types in the GaN:Eu system under study are an energy transfer to an impurity state outside of the *4f*-shell, and an Auger-type relaxation with free carriers. The PL lifetimes of GaN:Eu at room and low temperature indicate that there is only a minor quenching at room temperature under the conditions that the host material is not excited during $Eu^{3+}$ decay, about 80% of the $Eu^{3+}$ ions decay radiatively at room temperature[17,19]. In this case only the available thermally excited free carriers at elevated temperatures, and possibly nearby thermally excited states, contribute to the nonradiative deexcitation of $Eu^{3+}$ ions. During cw optical or electrical excitation, the concentration of free carriers and nearby excited states increases, and enhances the back-transfer, giving rise to a stronger quenching. In particular, an Auger-type energy transfer has been shown to effectively de-excite the $Eu^{3+}$ ions at high carrier densities[9,12].

Since the back-transfer rate is too small to explain the observed thermal quenching under pulsed excitation, the explanation must be sought in differences in the excitation efficiency of the $Eu^{3+}$ ions. When carriers are excited optically in the host material, there are several deactivation channels available which can reduce their population. Among non-radiative recombination processes is the trapping of carriers by Eu-related traps with an associated capture coefficient and concentration. These can then subsequently transfer their energy to the $Eu^{3+}$ ion, or otherwise de-excite by different processes, for example radiative, or non-radiative energy transfer to free or trapped carriers. For the minority site, the small thermal quenching and similarity between optical and electrical excitation, is indicating an efficient energy transfer from a bound exciton state[20]. The thermal quenching for this site can be fully explained by the lifetime shortening of the Eu-related PL, i.e. the energy transfer efficiency from the trap state to the $Eu^{3+}$ ions is ~100%. Contrarily, the majority site shows a much stronger quenching, indicating that the energy transfer efficiency from the associated state is much lower, and improves at low temperatures. Furthermore the majority site shows large differences between optical and electrical excitation. From earlier work it is known that the excitation of the majority site is likely related to donor-acceptor pair transitions[21]. The DAP is excited primarily by optical excitation, and upon recombination can transfer its energy to the Eu ion. Such a mechanism has also been observed for intentionally co-doped samples[22], which showed a correlation between (Mg-related) DAP emission and Eu emission, and increase for decreasing temperature.

The PL quantum efficiency has a maximum value of 0.285 for a photon fluence of $7 \times 10^{13}$ $cm^{-2}$. The total concentration of minority sites is estimated to be $10^{18}$ $cm^{-3}$, of which a fraction of $f * \sigma_{ex}$ = 0.1 is excited. For the majority site there is a concentration of $9 \times 10^{19}$ $cm^{-3}$ and the excited fraction is $8.4 \times 10^{-4}$. The total concentration of excited $Eu^{3+}$ ions will be $1.7 \times 10^{17}$ $cm^{-3}$ at RT and increases to $2.4 \times 10^{17}$ $cm^{-3}$ at 77 K. For this fluence, all of the efficient trapping centers will be occupied, which together with the QE of 0.46, sets an upper limit to the trap concentration $N_T = 2.0 \times 10^{17}$ $cm^{-3}$. Since the carbon concentration has been determined to be $1.2 \times 10^{16}$ $cm^{-3}$, we conclude it is more likely that the competing trap is related to the Ga vacancy. Furthermore, other competing channels, are also influencing the Eu emission efficiency. These can be radiative recombination of non-equilibrium carriers or trapping by other defects, but their contribution is much smaller due to low concentrations and/or trapping cross-sections.

A summary of the concentrations and effective excitation cross-sections found in this study is depicted in Table 1.

|  | Minority (N = $10^{18}$ $cm^{-3}$) | Majority (N = $5 \times 10^{19}$ $cm^{-3}$) |
| --- | --- | --- |
| PL | $1.6 \times 10^{-15}$ cm$^2$ | $1.2 \times 10^{-17}$ cm$^2$ |
| EL | $3.0 \times 10^{-15}$ cm$^2$ | $6.0 \times 10^{-18}$ cm$^2$ |

Due to the differences in excitation pathways by PL and EL, different values can be expected. In PL, light produces free carriers, which subsequently freely diffuse and get trapped, while under current injection the carriers are injected from a contact and flow through the sample. However, under a simple trapping-energy transfer mechanism, the ratio of the cross-sections shouldn't change for different Eu sites for both excitation conditions. In this case, however, the majority site shows a larger cross-section under optical excitation while the minority site shows a larger cross-section for electrical excitation. This is due to the fact that majority site has an associated energy transferring defect that can be excited optically as well, which has been observed by the possibility of sub-bandgap excitation[5]. Under above bandgap excitation, there is a contribution of non-equilibrium carrier trapping and direct optical excitation in the energy transfer to the majority site center.

**Conclusion**

The excitation efficiency of the majority and minority sites of Eu-doped for GaN have been studied under optical and electrical excitation. We have evaluated excitation cross-sections for both excitation conditions for the majority and minority incorporation sites. The minority site has an excitation cross-section which is 2 orders of magnitude larger than the majority site, and determines primarily the luminescence properties at low excitation fluences. Under these conditions the competition of efficient carrier trap, that we have related to unintentionally doped carbon, limits the external QE of Eu-related luminescence, however it still reaches 29 % at room temperature and 48% at 77 K. For large pump fluences the limited number of the minority sites, and the small excitation cross-section of the majority site, limit the quantum efficiency, as other non-radiative recombination channels will be more effective in decreasing the excited carrier concentration. These results show the potential of Eu:GaN as efficient light emitters, especially in devices operating with low-current densities, like micro-LEDs.

1. Zhu, W. *et al.* High-Power Eu-Doped GaN Red LED Based on a Multilayer Structure Grown at Lower Temperatures by Organometallic Vapor Phase Epitaxy. *MRS Adv.* **2**, 159–164 (2017).
2. Mitchell, B., Dierolf, V., Gregorkiewicz, T. & Fujiwara, Y. Perspective: Toward efficient GaN-based red light emitting diodes using europium doping. *J. Appl. Phys.* **123**, 160901 (2018).
3. Fleischman, Z. *et al.* Excitation pathways and efficiency of Eu ions in GaN by site-selective spectroscopy. *Appl. Phys. B* **97**, 607–618 (2009).
4. Woodward, N., Nishikawa, A., Fujiwara, Y. & Dierolf, V. Site and sample dependent electron–phonon coupling of Eu ions in epitaxial-grown GaN layers. *Opt. Mater. (Amst).* **33**,

1050–1054 (2011).

5. Fujiwara, Y. & Dierolf, V. Present understanding of Eu luminescent centers in Eu-doped GaN grown by organometallic vapor phase epitaxy. *Jpn. J. Appl. Phys.* **53**, (2014).

6. Roqan, I. S. *et al.* Identification of the prime optical center in GaN : Eu 3 +. *Phys. Rev. B* **81**, 085209 (2010).

7. Wakamatsu, R. *et al.* Luminescence properties of Eu-doped GaN grown on GaN substrate. *Jpn. J. Appl. Phys.* **52**, 08JM03 (2013).

8. Mitchell, B. *et al.* Charge state of vacancy defects in Eu-doped GaN. *Phys. Rev. B* **96**, 064308 (2017).

9. Zhu, W. *et al.* Re-Excitation of Trivalent Europium Ions Doped into Gallium Nitride Revealed through Photoluminescence under Pulsed Laser Excitation. *ACS Photonics* **5**, 875–880 (2018).

10. Zhu, W. *et al.* Enhanced photo/electroluminescence properties of Eu-doped GaN through optimization of the growth temperature and Eu related defect environment. *APL Mater.* **4**, 056103 (2016).

11. Nishikawa, A., Kawasaki, T., Furukawa, N., Terai, Y. & Fujiwara, Y. Room-Temperature Red Emission from a p-Type/Europium-Doped/n-Type Gallium Nitride Light-Emitting Diode under Current Injection. *Appl. Phys. Express* **2**, 071004 (2009).

12. Mitchell, B. *et al.* Color-Tunablility in GaN LEDs Based on Atomic Emission Manipulation under Current Injection. *ACS Photonics* **6**, 1153–1161 (2019).

13. Honda, U., Yamada, Y., Tokuda, Y. & Shiojima, K. Deep levels in n-GaN Doped with Carbon Studied by Deep Level and Minority Carrier Transient Spectroscopies. *Jpn. J. Appl. Phys.* **51**, 04DF04 (2012).

14. Auret, F. D. *et al.* Electrical characterisation of hole traps in n-type GaN. *Phys. status solidi* **201**, 2271–2276 (2004).

15. Tokuda, Y. *et al.* Hole traps in n-GaN detected by minority carrier transient spectroscopy. *Phys. Status Solidi Curr. Top. Solid State Phys.* **8**, 2239–2241 (2011).

16. Armstrong, A. *et al.* Impact of deep levels on the electrical conductivity and luminescence of gallium nitride codoped with carbon and silicon. *J. Appl. Phys.* **98**, 053704 (2005).

17. de Boer, W. D. A. M. *et al.* Optical excitation and external photoluminescence quantum efficiency of $Eu^{3+}$ in GaN. *Sci. Rep.* **4**, 5235 (2014).

18. Boyn, R. 4f–4f Luminescence of Rare-Earth Centers in II–VI Compounds. *Phys. status solidi* **148**, 11–47 (1988).

19. Wakamatsu, R., Lee, D., Koizumi, A., Dierolf, V. & Fujiwara, Y. Luminescence properties of Eu-doped GaN under resonant excitation and quantitative evaluation of luminescent sites. *J. Appl. Phys.* **114**, 043501 (2013).

20. Zimmermann, H. & Boyn, R. Donor-Type Tm Centres in ZnS Crystals. *Phys. status solidi* **139**, 533–545 (1987).

21. Mitchell, B. *et al.* The role of donor-acceptor pairs in the excitation of Eu-ions in GaN:Eu epitaxial layers. *J. Appl. Phys.* **115**, 204501 (2014).
22. Singh, A. K. *et al.* Eu–Mg defects and donor–acceptor pairs in GaN: photodissociation and the excitation transfer problem. *J. Phys. D. Appl. Phys.* **51**, 065106 (2018).

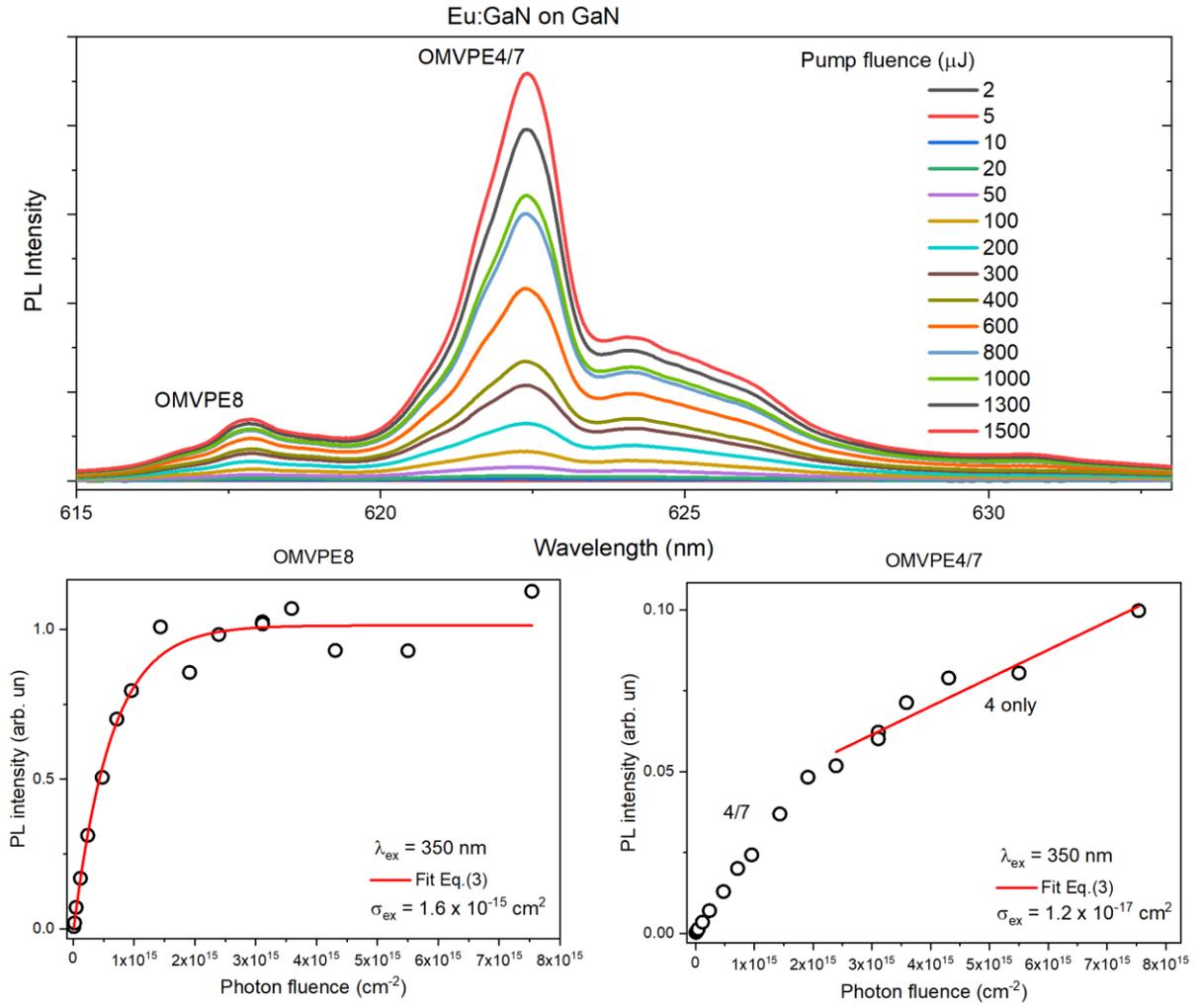

*Figure 1 Top: PL spectra for an excitation wavelength of 350 nm for various pump fluences. Bottom: Intensity of the peaks related to OMVPE8 (left) and OMVPE4/7 (right) as obtained by spectral deconvolution of the PL spectra. The vertical axis indicate the relative fraction of excited centers. The red lines show fits of Eq. (3).*

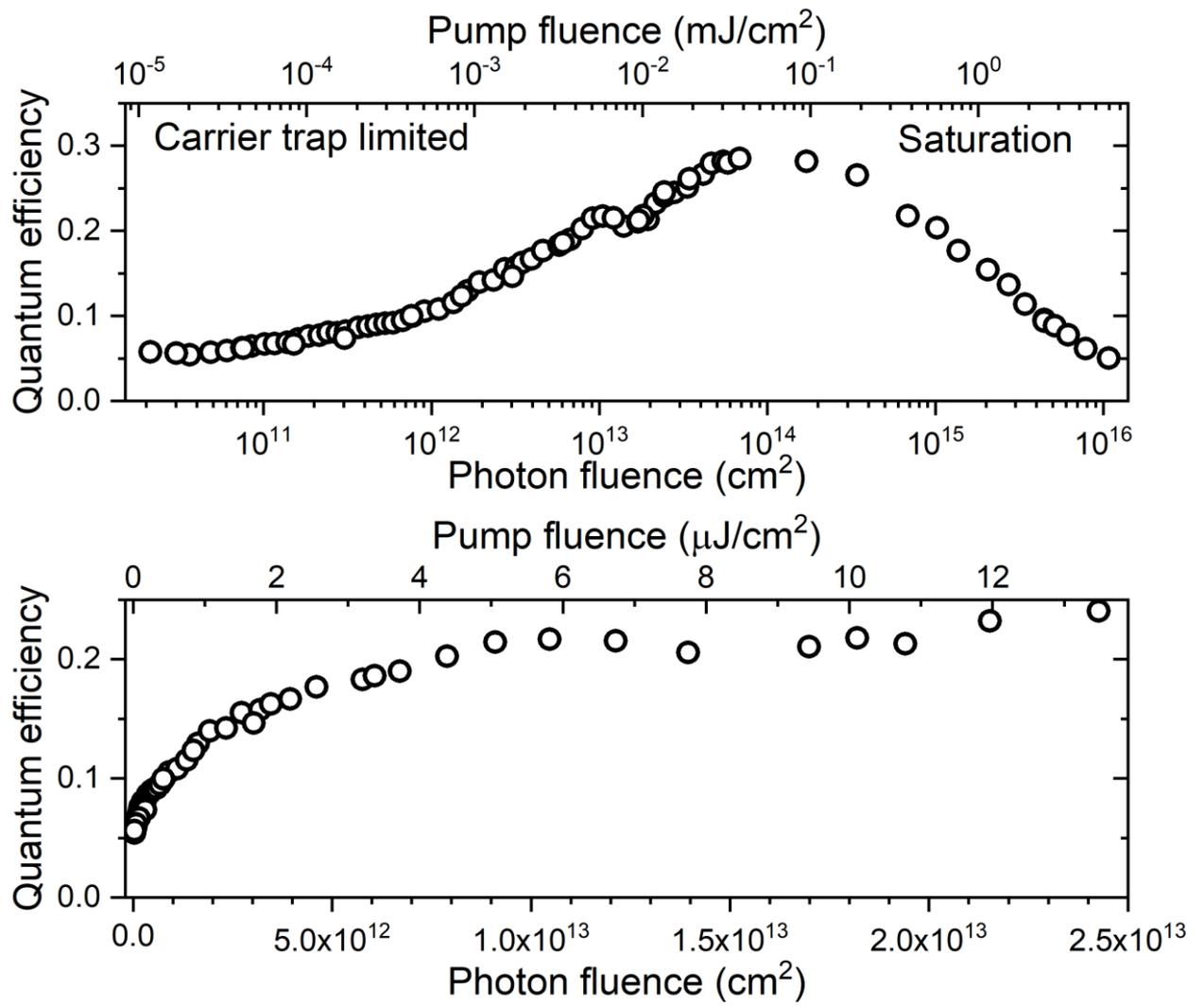

*Figure 2 Top: Dependence of the external QE of Eu-related emission on the pump fluence over 6 orders of magnitude. Bottom: linear plot for the low pump fluence region.*

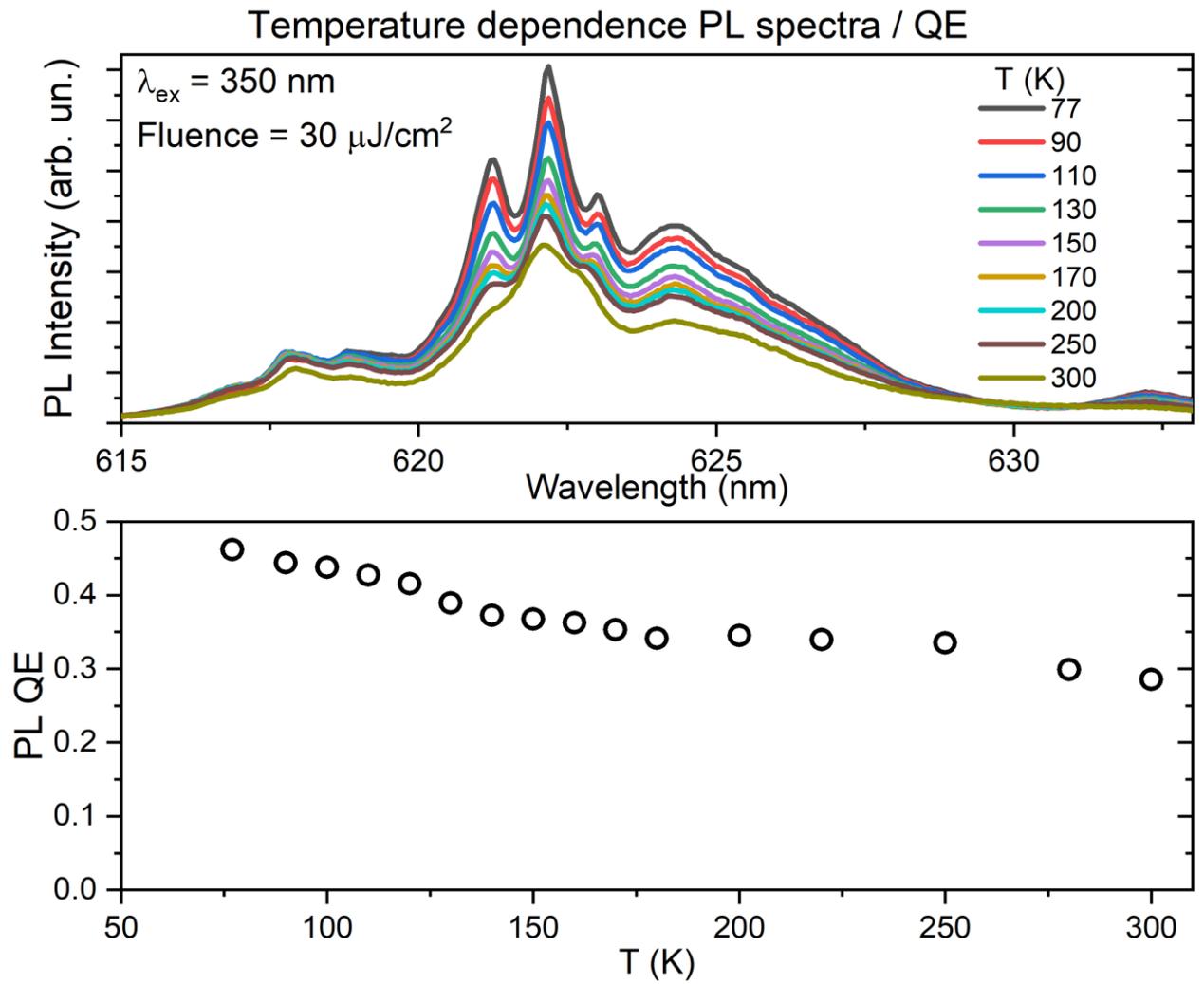

*Figure 3 Temperature dependence of the PL spectrum (top) and PL QE (bottom) of the Eu-related emission.*

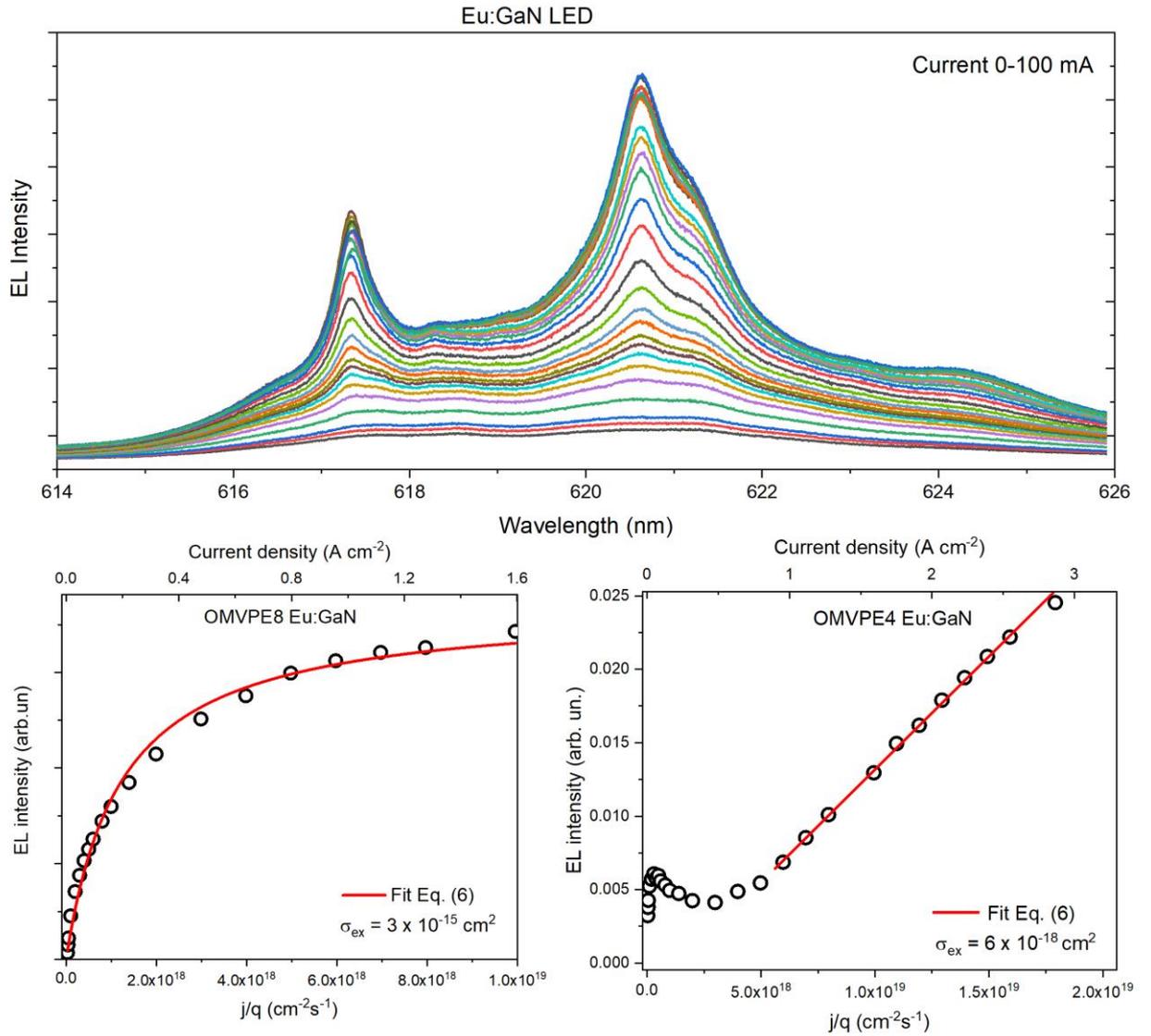

*Figure 4 Top: EL spectra for various injection currents up to 100 mA. Bottom: Intensity of the peak related to OMVPE8 (left) and OMVPE4 (right) as obtained by spectral deconvolution of the EL spectra and the following the procedure in the main text. The vertical axis indicate the relative fraction of excited centers. The red lines show fits of Eq. (6).*